\documentclass[graybox,envcountchap,sectrefs]{svmono}

\usepackage{type1cm}         

\usepackage{natbib}
\usepackage{makeidx}         
\usepackage{graphicx}        
\usepackage{multicol}        
\usepackage[bottom]{footmisc}

\usepackage{newtxtext}       %
\usepackage{newtxmath}       
\usepackage[normalem]{ulem}
\useunder{\uline}{\ul}{}
\usepackage{authblk}

\makeindex             

\begin{document}

\author[1]{Gowri Sankar Ramachandran}
\author[1]{Sidra Malik}
\author[1]{Shantanu Pal}
\author[1]{Ali Dorri}
\author[2]{Volkan Dedeoglu}
\author[3]{Salil Kanhere}
\author[1]{Raja Jurdak} 
\affil[1]{Queensland University of Technology, Brisbane, Australia.
Email: \{g.ramachandran,s.malik, shantanu.pal, ali.dorri, r.jurdak\}@qut.edu.edu}
\affil[2]{CSIRO, Brisbane, Australia.
Email: volkan.dedeoglu@data61.csiro.au}
\affil[3]{University of New South Wales, Sydney, Australia.
Email: salil.kanhere@unsw.edu.au}
\title{Blockchain in Supply Chain: Opportunities and Design Considerations}
\maketitle





%
%
%
\chapter{Blockchain in Supply Chain: Opportunities and Design Considerations}
\label{intro} 

\abstract{Supply chain applications operate in a multi-stakeholder setting, demanding trust, provenance, and transparency. Blockchain technology provides mechanisms to establish a decentralized infrastructure involving multiple stakeholders. Such mechanisms make the blockchain technology ideal for multi-stakeholder supply chain applications. This chapter introduces the characteristics and requirements of the supply chain and explains how blockchain technology can meet the demands of supply chain applications. In particular, this chapter discusses how data and trust management can be established using blockchain technology. The importance of scalability and interoperability in a blockchain-based supply chain is highlighted to help the stakeholders make an informed decision. The chapter concludes by underscoring the design challenges and open opportunities in the blockchain-based supply chain domain.}

\section{Introduction}
\label{introduction}
\label{sec:1}
Supply chains drive many industries in multiple domains, including agriculture [\cite{kamilaris2019rise,kamble2020modeling}], electronics[\cite{lee2017blockchain,xu2019electronics}], and manufacturing[\cite{abeyratne2016blockchain,kurpjuweit2021blockchain}]. Enterprises procure materials and services from different stakeholders to build, store and deliver products to end-consumers. Information about the product, referred to as ``data," gets exchanged between manufacturers, shipment companies, auditors, regulators, and retailers in this product pipeline. Businesses rely on data to understand the status of the supply chain while making an informed decision about future demands and optimising their operations. Besides, the data helps the companies manage the regulatory compliance and auditing processes.

To sum up, the information about the supply chain, in the form of digital data, can enhance the efficiency of the supply chain. However, the processes followed by the supply chain entities must be trustworthy to attain meaningful insights. Blockchain technology offers mechanisms to enhance trust. 

Bitcoin[\cite{nakamoto2019bitcoin}], a blockchain platform, introduced an innovative architecture for creating, managing, and sharing a digital currency without involving a centralized intermediary such as a bank. It offers decentralization, immutability, and transparency through a clever combination of a consensus algorithm, cryptographic primitives, and a distributed ledger. After the arrival of Bitcoin, many blockchain platforms, including Ethereum[\cite{wood2014ethereum}], Tendermint[\cite{buchman2016tendermint}], and EOS[\cite{grigg2017eos}], entered the market with similar properties and support for cryptocurrencies. However, the properties of Bitcoin and other blockchain platforms that followed it seem extremely useful for applications beyond cryptocurrency, that require decentralization, immutability, and transparency.

Supply chain applications' natural characteristics include its involvement of multiple stakeholders in the form of producers, retailers, auditors, shipment companies, and in some cases, the end consumers. Having a transparent operational infrastructure driven by a decentralized architecture combined with an immutable ledger provide enormous business and practical advantages to the stakeholders. Here, the integration of blockchain technology into the supply chain requires IT infrastructure. Many of the supply chain companies already employ an IT infrastructure to manage their business processes digitally[\cite{wu2006impact}]. Integrating blockchain technology into the existing supply chain management infrastructure allows the supply chain participants to gain business advantages by improving traceability and transparency, automating processes for purchasing and payments, reducing conflicts and errors, improving regulatory compliance and cross-border transactions, and protecting the supply chain against counterfeiting through an immutable and tamper-proof distributed ledger. Therefore, the benefits of combining blockchain technology with the supply chain application are clear. Still, it is crucial to understand how blockchain technology can be merged reliably and seamlessly with supply chain networks to gain maximum advantages.

This chapter discusses blockchain applications in the supply chain and highlights the characteristics and requirements of supply chain applications. We also explain how supply chain stakeholders can share data and manage trust using blockchain technology. Besides, we also focus on the need for scalable blockchain technology to deal with the high transaction throughput and latency demands of real-world supply chain applications. The section on authentication and access control discusses methods to set up a trusted and secure supply chain infrastructure. The interoperability challenges of supply chain systems are also discussed to help companies understand the existing challenges and potential approaches. We conclude the chapter with an overview of design considerations and open challenges.

\section{Characteristics and Requirements of Supply Chain Applications}
\label{characteristics-requirements}

\begin{figure}
    \centering
    \includegraphics[scale=0.4]{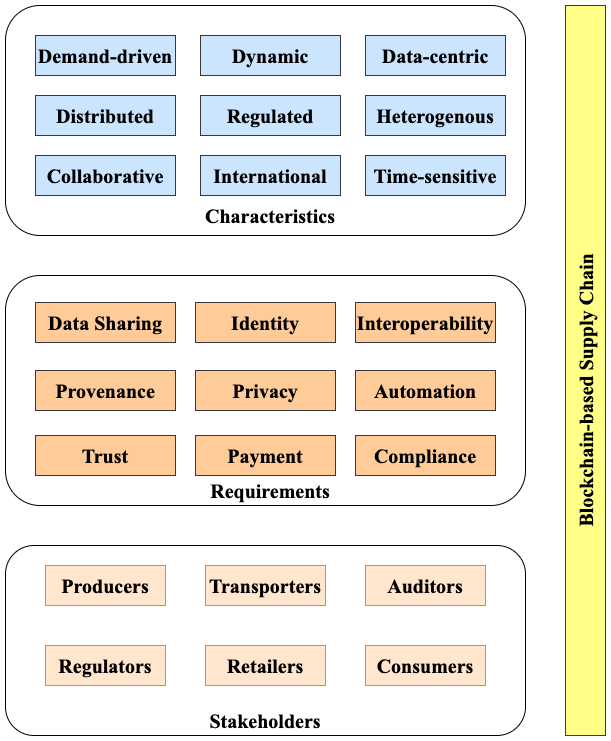}
    \caption{Characteristics and Requirements of Blockchain-based Supply Chain.}
    \label{fig:overview}
\end{figure}

A supply chain can be defined as the end-to-end process of producing and delivering goods and services from the acquisition of raw materials to the delivery of final products to the end-consumers. As depicted in Figure~\ref{fig:overview}, a typical supply chain involves multiple stakeholders and has many characteristics and requirements. This section provides an overview of these characteristics and common requirements together with the main challenges facing today's supply chain.

\subsection{Characteristics of Supply Chain Applications and Networks}
In this section, we will review the characteristics of supply chain applications.
\begin{itemize}
    \item \textbf{Collaborative:} Supply chain applications naturally involve multiple stakeholders in the form of producers, transporters, auditors, retailers, and regulators as shown in Figure~\ref{fig:overview}. Each of these stakeholders must collaborate with each other to operate the supply chain in the process of delivering goods and services to end consumers while gaining financial incentives. However, collaboration becomes challenging when the level of trust is low and there is no effective mechanism to manage and share supply chain data transparently among stakeholders. 
    
    \item \textbf{International:} Globalization of production and trade has given rise to geographic dispersion of the supply chain, where the stakeholders may operate from different legal jurisdictions and countries. In a global supply chain, raw materials may be acquired from suppliers, processed by manufacturers and the final products may be delivered to consumers at geographically dispersed locations introducing new risks and challenges in the flow of goods and services. Note that the stakeholders may have to comply with the regulations of one or more jurisdictions in this setting. 
    
    \item \textbf{Time-sensitive:} Supply chain processes are time-sensitive. Stakeholders rely on the timely flow of goods and information to make operational decisions. Any delay in the flow of goods and information may cause supply chain disruptions in the form of inefficiencies, quality degradation of products, lower service quality, market loss, and reduced gains.
    
    \item \textbf{Distributed:} In a supply chain, the information technology (IT) and operation technology (OT) infrastructure that manages the flow of goods, documents, certifications, and other information is distributed among multiple stakeholders. These infrastructures are typically connected through enterprise systems and data sharing frameworks. 
    
    \item \textbf{Regulated:} Supply chains are subject to national and international (import and export) regulations. Stakeholders must comply with the laws and regulations of their jurisdiction. In some cases, they may have to comply with jurisdictions associated with both the producer and consumer. Such complexities make regulatory compliance challenging, in particular for supply chains operating in multiple jurisdictions. 
    
    \item \textbf{Heterogenous:} Supply chains involve multiple stakeholders using heterogeneous data standards and disparate systems. Note that different standards may introduce compatibility issues, which could make collaboration challenging and difficult in a supply chain application. 
    
    \item \textbf{Demand-driven:} The dynamic nature of consumer behaviour requires supply chains to be resilient against shifting consumer demand. Demand-driven supply chain management focuses on the demand signals and forecasts for the planning and operation of processes to optimize the delivery of goods and services to consumers. Building a supply chain that is able to respond to changing consumer demand may require supply chain stakeholders to invest in technology to collect, share, and react to real-time demand data.
    \item \textbf{Dynamic and Unpredictable:} Supply chains are dynamic systems due to ongoing changes in stakeholders, offered services and products, regulations, market dynamics, and technological advancements. For example, when a ship accidentally blocked the Suez Canal in 2021, many ships could not cross the Suez Canal for up to six days, which disrupted the supply chain massively. To respond to such unpredictable real-world events, flexible and agile supply chain mechanisms are required. 
    
    \item \textbf{Data-centric:} For improved performance, supply chain operations should be based on data-driven decision making, which relies on real-time, accurate, and actionable data flow. Digitisation of supply chains, data provenance and trust mechanisms, big data analytics, and effective data sharing systems help stakeholders make informed business decisions to enhance the efficiency of their supply chain. 
\end{itemize}

\subsection{Requirements of Supply Chain}
In this section, we will review the common requirements of supply chain applications.
\begin{itemize}
    \item \textbf{Trust:} Trust is a pivotal factor in enabling collaboration in supply chains. Since supply chains are complex constructs, establishing trust among the stakeholders and the systems used is a challenging requirement for the integrity of supply chains. 
    \item \textbf{Payment:} In a supply chain transaction, a seller provides  goods and services to a buyer and receives a payment in return. Payment systems enabling business-to-business payments without relying on third parties, accelerating the flow of funds, reducing the transaction costs, and lowering the risk of financial frauds improve the sustainability and efficiency of supply chains. 
    \item \textbf{Compliance:} In a supply chain, stakeholders need to comply with various regulations, standards, and policies. Non-compliance may result in financial loss due to legal fines, potential loss of market, loss of non-compliant assets, and supply chain disruptions as well as brand and reputation damage for companies. Thus, supply chain stakeholders need to implement mechanisms to ensure regulatory compliance. Recently, regulatory technology (RegTech) has been proposed as a potential solution to address the complexities of compliance through digitization and automation of the compliance process.
    \item \textbf{Provenance:} Due to the rising consumer demand on the information related to the origin of products, the processes for manufacturing and production, transfer of custody and ownership, provenance mechanisms become critical. In a supply chain, provenance improves consumer confidence and trust in the authenticity of the products and accountability among stakeholders.
    \item \textbf{Privacy:} To improve the operational efficiency of supply chains, the stakeholders are required to share their data between stakeholders in a supply chain network. However, supply chain data includes commercially sensitive information. Companies may feel that revealing such data may cause a business to lose its competitive advantage. Thus, privacy preservation mechanisms are required for enabling data sharing while protecting the sensitive business data.
    \item \textbf{Automation:} In a typical supply chain, there are many time-consuming, error-prone and repetitive processes that can be automated, such as processing orders and payments, inventory management, transportation arrangements, and manufacturing workflows. Using emerging technologies such as artificial intelligence, robotics, IoT, and big data, supply chain processes can be automated to enhance the efficiency by lowering operational costs, reducing manual labor, improving inventory management, and increasing productivity and accuracy while ensuring continuous compliance to regulations.
    \item \textbf{Data sharing:} Data sharing improves coordination and trust among stakeholders, efficiency of production, quality of products and services, while reducing operational costs and risk of non-compliance. Furthermore, establishing end-to-end provenance requires supply chain stakeholders to share their data with each other and the end-consumers. Therefore, trusted and effective mechanisms are needed to enable data sharing among distributed and heterogeneous stakeholders. Lack of incentives may constitute a barrier for data sharing. Thus, incentive mechanisms can be utilized to encourage stakeholders to share their data. 
    \item \textbf{Identity management:} Supply chains are distributed and complex networks, where the verification of the identities and credentials of interacting stakeholders and the digital identities for the products and devices involved is crucial for the cooperation among stakeholders and the operation of supply chains. However, stakeholders may not be willing to disclose or share the data related to supply chain identities due to business sensitivities and to uphold competitive advantage. As a consequence, supply chains may suffer from lack of transparency, which becomes a bigger problem as the supply chains grow. Traditional centralised and federated identity management approaches establish trust between stakeholders relying on centralised entities or platforms, which introduces data ownership and security problems. Thus, decentralised mechanisms are required to manage supply chain identities, where entities control their data shared with other stakeholders without relying on third parties while enabling authorities to verify the identities and credentials of the stakeholders.
    \item \textbf{Interoperability:} Due to the interconnected and interdependent nature of supply chain networks, interoperability is required to facilitate collaboration among the stakeholders. While interoperability can be achieved through standards, procedures, and inter-organisational protocols at the service and business levels, digital interoperability requires mechanisms and protocols for application interfaces, integration services, and data communications to enable reliable and seamless data sharing among the distributed and heterogeneous supply chain systems. 
\end{itemize}
Supply chain applications could become trustworthy, transparent, and efficient if the above requirements are fulfilled. While existing supply chain systems fulfill these requirements through a combination of manual and automated processes, they broadly rely on systems managed by a single organization. Relying on a single organization introduces a single point of failure, wherein the entity may misbehave or tamper with the supply chain information. In the rest of this chapter, we will introduce blockchain technology and its application in supply chain to fulfill some of these requirements.

\section{Overview of Blockchain Technology}
\label{sec:blockchain}
The characteristics and requirements of the supply chain applications demand solutions that can operate in a distributed setting while offering trust, security, data sharing, payment, and transparency guarantees. Supply chain operators can set up an infrastructure to share information with other stakeholders using a custom-built and centrally managed platform. However, such solutions are susceptible to central points of failure, wherein the organization that runs the platform has complete control of the infrastructure. In a supply chain network, a centrally managed infrastructure could be compromised by a malicious stakeholder, resulting in incorrect dispersion of information. The stakeholders may not be willing to reveal essential traces and logs in the case of a dispute or unexpected events due to their error. Note that the organizations would like to ensure that their reputation is not damaged when they mismanage the supply chain. Therefore, the centrally managed infrastructure is not guaranteed to provide complete transparency to the stakeholders in the network.

\subsection{Introduction to BitCoin}

Blockchain technology provides support for distributed and decentralized infrastructure. At its core, it consists of a consensus algorithm, cryptographic protocols, and immutable storage. The first peer-to-peer electronic cash system, BitCoin[~\cite{nakamoto2019bitcoin}], introduced a blockchain that cleverly distributes the data validation process to the computation nodes in the public network. Following the BitCoin protocol, any computationally capable node can create blocks that make up the blockchain using a consensus mechanism called Proof-of-Work (PoW). PoW lets the public BitCoin nodes solve a computationally intensive cryptographic puzzle, which demands significant computation resources. On average, the nodes in the network require 10 minutes to solve the puzzle. When a node solves the puzzle, it is selected as a winner for that round. The winning node gets to verify the new transactions and propose the new block. Subsequently, all the nodes in the network receive the newly proposed block, verify them, and append to their local copy of the blockchain ledger. 

The ledger used by the blockchain technology offers immutability support, which means, once the data is written to the blockchain, it cannot be modified. When a malicious entity tries to alter the data on the blockchain, he gets to modify only his local copy of the ledger. Besides, other nodes in the network will not accept such modifications because the modified version will fail the integrity check when it is compared with the original blockchain ledger maintained by the majority of the nodes in the network. Note that hundreds to thousands of nodes in the network keep a copy of the ledger. It is nearly impossible to update the ledger on all of those nodes. Therefore, the write-once property of blockchain technology is one of the powerful features for applications that require trust and transparency.

Blockchain technology also enables the nodes in the network to maintain some level of anonymity through the use of public-key cryptography. Users submit transactions by using their public key and a signature, which forms the basis of their identity. When a transaction is submitted by the user, for example, to spend a BitCoin, the user digitally signs the transaction using her private key and then presents the signature and a public key to the network. Anyone in the BitCoin network can verify the authenticity of the signature using the public key. Note that the user need not share his private key. 

In summary, the BitCoin platform, which started the blockchain revolution, uses PoW to create and manage blocks without a central intermediary securely. Besides, public-key cryptography protects users' privacy while allowing them to exchange digital currencies in a trusted manner. Lastly, the immutable ledger maintains a distributed and trusted record of all the transactions to provide complete transparency. 

\subsection{Other Blockchain Platforms}

Following the introduction of BitCoin, many blockchain platforms entered the industry, offering immutability, transparency, and trust guarantees. Notably, Ethereum[\cite{wood2014ethereum}] is one of the most popular blockchain platforms, after BitCoin. While Ethereum's architecture closely resembles BitCoin, including its use of PoW \footnote{Note that the Ethereum community is developing a new Proof-of-Stake consensus protocol, which may soon replace the energy-inefficient PoW protocol.}, public-key cryptography, and immutable ledger, it introduced smart contracts as a new feature. Smart contracts allow the users to run computations within the blockchain platform to enable automation. For example, users can code up the supply chain events in the smart contract and let the smart contract automatically transfer payments to the relevant stakeholders when the event associated with the product delivery event triggers the smart contract on the blockchain. This functionality of Ethereum resulted in many innovations in the finance, supply chain, and banking domains.

When BitCoin and Ethereum became popular blockchain technology for decentralized and trusted computing, a new blockchain deployment model entered the industry. Platforms such as BitCoin and Ethereum use an open and public network model; wherein anyone can join the blockchain network to participate in the creation and management processes. A new permissioned deployment model was introduced to let enterprises set up a private blockchain ledger to manage multi-stakeholder transactions. Hyperledger Fabric[\cite{androulaki2018hyperledger}] is one of the most popular permissioned blockchain platforms. It is a lightweight blockchain platform with a trusted and immutable ledger. In Hyperledger Fabric, the consensus process is managed by an orderer, which verifies the integrity of the transactions. And a set of nodes are assigned to manage the ledger for the entire network. While a permissioned blockchain network follows a private setup, it still offers trust, immutability, and transparency properties. However, users must prove their identity and follow an on-boarding process to become part of a permissioned blockchain network. In contrast, the nodes' identities are unknown in the public blockchain platforms such as BitCoin and Ethereum.

To sum up, blockchain technology provides a decentralized infrastructure with support for trust, transparency, immutability, payment, and computation (through smart contract). Supply chain applications could leverage these functionalities to fulfill the requirements listed in Section~\ref{characteristics-requirements}. The following section will review the popular blockchain-based supply chain applications.

\section{Real-world Applications of Blockchain in Supply Chain}
\label{sec:apps}
Many blockchain-based supply chain applications have been discussed in the literature[\cite{https://doi.org/10.1111/jbl.12240,lee2017blockchain,blockchain2019ibm,jensen2019tradelens}]. In this section, we will review two of the most popular and active real-world blockchain-based supply chain applications.

\subsection{IBM FoodTrust}
IBM developed a food supply chain system using blockchain technology to address the food inefficiency problems. Improper management of the food supply chain and the lack of visibility into the food production processes lead to high food waste, increased carbon footprint, health issues due to contamination, and high pricing of goods. To overcome these inefficiencies, IBM created FoodTrust[\cite{blockchain2019ibm}], which connects industries, farmers, and other stakeholders in the food and agriculture industry through a blockchain-based system. The key features of FoodTrust include Insights, Trace, and Documents modules, which are discussed below:
\begin{itemize}
    \item \textbf{Insights:} This module uses blockchain technology in combination with the Internet of Things (IoT) to provide visibility to the supply chain. With this module, the supply chain stakeholders can understand how fresh a food product is in near real-time.
    \item \textbf{Trace:} This module provides end-to-end traceability for the supply chain stakeholders. It allows the members to trace the location and the status of the food products in a secure and trusted manner.
    \item \textbf{Documents:} This module helps the members in the supply chain network securely manage certificates and other digital documents. Regulators and auditors can ensure compliance through this module. 
\end{itemize}

Through these modules, IBM FoodTrust allows the stakeholders to enhance their reputation in the market. IBM manages this project through the Platform-as-a-Service model. Organizations interested in providing transparency and provenance can join IBM's FoodTrust to manage their supply chain.


\subsection{TradeLens}
\begin{figure}
    \centering
    \includegraphics[scale=0.3]{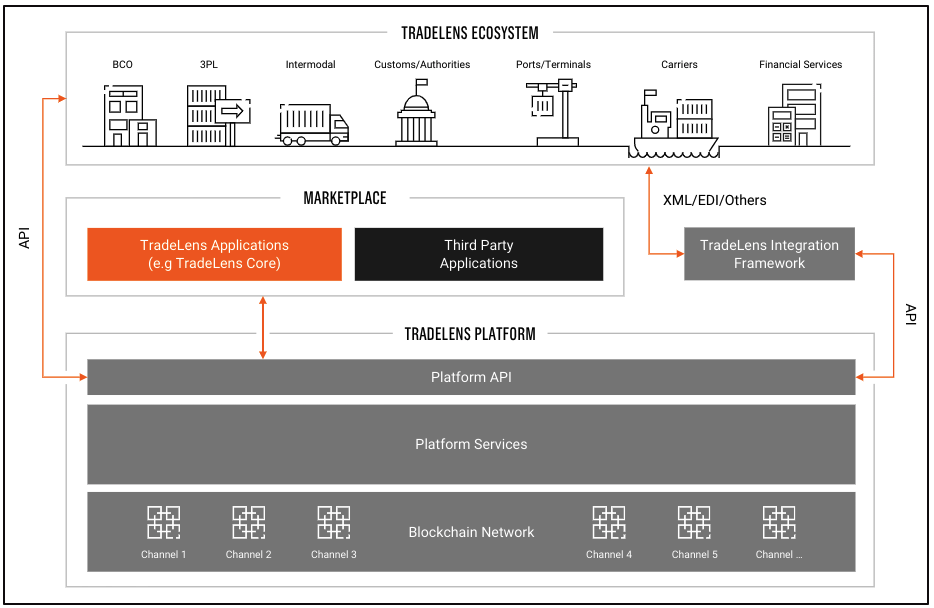}
    \caption{Architecture of TradeLens Platform [\cite{jensen2019tradelens}].}
    \label{fig:tradelens}
\end{figure}

TradeLens[\cite{jensen2019tradelens}] is a blockchain-based platform for a global supply chain. IBM and GTD Solution Inc jointly develop the platform. In particular, TradeLens focuses on shipping and transportation processes to cater to the demands of the logistics operators. As of June 2021, the TradeLens platform handled more than 2 billion events, millions of documents, and more than 40 million containers.  It is a permissioned platform with support for privacy preservation, immutable storage, enterprise-grade security, and standards-compatible. Figure~\ref{fig:tradelens} provides an overview of TradeLens' architecture.

TradeLens provides Open APIs to let organizations interact with a platform. The supply chain operators, including port authorities, shipping companies, and consignment owners, can easily leverage the TradeLens platform through interoperable and standard-based APIs. Stakeholders with legacy enterprise and IT systems have to invoke the REST APIs, which securely register the supply chain data in the blockchain.

\subsection{Other applications}
In addition to the platforms managed by IBM, a few other applications also employ blockchain technology in the supply chain. EverLedger~\cite{underwood2016blockchain} tracks the journey of diamond using blockchain technology. Another notable application is BeefLedger~\cite{cao2020beefledger}, which tracks the beef supplies using Ethereum and Proof-of-Authority consensus. Besides, the healthcare domain also explores blockchain technology to track medicines and manage patient's healthcare data~\cite{radanovic2018opportunities,9474701}.

\section{Blockchain-based Data sharing for Supply Chain}
\label{bc-data-sharing}
Supply chain applications constantly share data with other organizations in the application network. This information includes the product's status, invoices, regulatory requirements, demands, and customer feedback. A combination of digital and manual processes is needed to collate the information and then forward them to the necessary stakeholders through either their ERP systems or email messages. 

\begin{figure}
    \centering
    \includegraphics[scale=0.3]{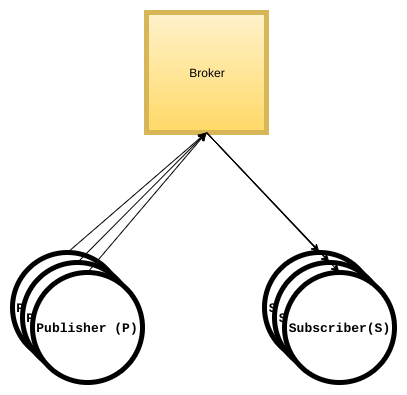}
    \caption{Overview of Publish-Subscribe Broker.}
    \label{fig:pub-sub}
\end{figure}

In a multi-stakeholder environment, each organization may have a digital infrastructure and an operational procedure to exchange information relevant to their business. In some cases, a third-party data-sharing platform is leveraged by all the stakeholders for data sharing. Businesses that rely on a third-party data sharing platform hand over the management responsibility to an organization that is not necessarily part of the supply chain network. In such circumstances, all the organizations in the supply chain networks must trust the third party. Such an architecture could fulfill the stakeholders' business and data sharing demands, but it assumes that the third party is honest and trustworthy.

Blockchain technology offers a decentralized, distributed, and transparent platform for application developers. Supply chain applications could leverage blockchain technology in a multi-stakeholder environment. Using blockchain technology, each organization that is part of the supply chain network could participate in the data sharing process while having access to the transparent and immutable ledger. In the rest of this section, we will review Trinity, a distributed publish-subscribe broker with blockchain-based immutability.


\begin{figure}
    \centering
    \includegraphics[scale=0.3]{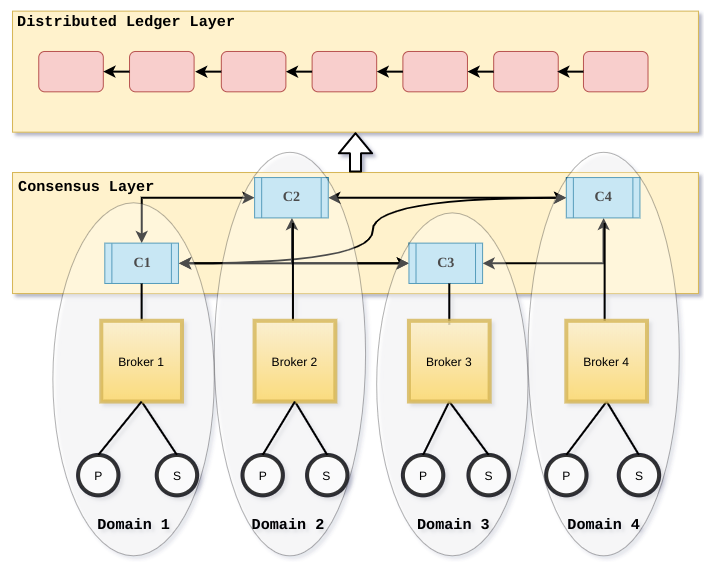}
    \caption{The Architecture of Trinity: A Distributed Publish-Subscribe Broker[\cite{8751388}].}
    \label{fig:trinity}
\end{figure}

Multiple data sharing modalities have been considered for supply chain applications including publish-subscribe messaging model. Because of its lightweight design, resource efficiency, and scalability, the publish-subscribe messaging model is widely used in supply chain applications, which is evident from the list of customers including McDonald's that use PubNub, a publish-subscribe service provider. In a nutshell, the pub-sub messaging model connects the data providers (denoted as publishers) with the data consumers (denoted as subscribers) through a topic-based interaction model, which is shown in Figure~\ref{fig:pub-sub}. For example, a shipment dispatched from a manufacturer's hub could be labelled as ``stakeholderA/countryX/shipmentid6576" and the data associated with this shipment could be accessed by all the relevant stakeholders by simply subscribing to topic ``stakeholderA/countryX/shipmentid6576". Following this design philosophy, a topic can be created for each shipment to let the stakeholders both send (publish) and receive (subscribe) data. 

Despite its advantages, the publish-subscribe system follows a centralized architecture, wherein the data publishers and subscribers interact via a centralized broker. Here, the broker receives the data from publishers and routes it to the relevant subscribers. In this architecture, it is clear that the broker is critical for data sharing. Therefore, the organisation that runs the broker must act honestly and it should not tamper with the data. Recall that the centralized solutions are prone to Byzantine failures. Trinity[~\cite{8751388}] is a Byzantine fault-tolerant and distributed publish-subscribe broker, that can operate in a multi-stakeholder setting while preventing Byzantine failures and without breaking the fundamental interaction behaviour of publish-subscribe messaging model. Figure~\ref{fig:trinity} shows the architecture of Trinity[~\cite{8751388}]. 

Trinity is a distributed Byzantine fault-tolerant pub-sub broker[~\cite{8751388}]. It is created for multi-stakeholder applications, in which each stakeholder is assumed to be operating a broker and a consensus node within their domain to serve its local publishers and subscribers. All the domains are connected through a consensus layer, which is responsible for validating published messages. Each message must be approved by more than one-third of the consensus nodes, before it is sent to the subscribers via a broker. Trinity ensures that when a subscriber in one of the domains receives a message, all the other subscribers in a non-faulty domain also receive the same message, provided more than one-thirds of the domains are non-faulty. A set of safety and liveness properties are presented here, if any reader is interested in understanding the trust and safety guarantees in detail [\cite{8751388}].

\begin{figure}
    \centering
    \includegraphics[scale=0.27]{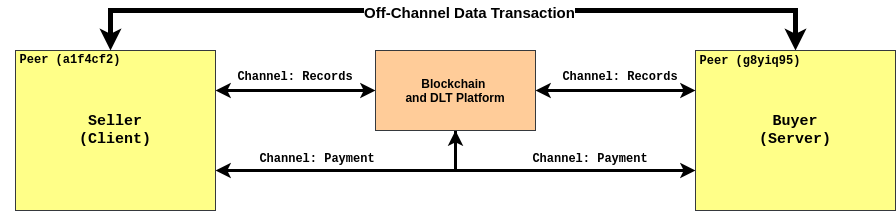}
    \caption{Overview of Streaming Data Payment Protocol[~\cite{radhakrishnan2019sdpp}].}
    \label{fig:sdpp}
\end{figure}

\textbf{Streaming Data Payment Protocol:} Streaming data payment protocol (SDPP) is another blockchain-based data sharing platform, which incentivizes the data producers. In a supply chain setting, SDPP can be used to handle payments whenever the data associated with the supply chain is shared among stakeholders. It enables a data provider and data consumer to easily connect and transact with each other using micropayments for streaming data[~\cite{sdpp,radhakrishnan2019sdpp}]. SDPP is a peer-to-peer data sharing protocol. Its design carefully separates out three key components: the off-chain data communication channel (which is operated as a traditional Internet client-server application-layer protocol, atop TCP), a payment channel (implemented using a cryptocurrency protocol), and a records medium (implemented using a distributed ledger technology), as shown in Figure~\ref{fig:sdpp}.

Alternatively, applications can interface with third-party blockchain-as-a-service platforms such as IBM's FoodTrust and TradeLens, discussed in Section~\ref{sec:apps}. Note that the supply chain systems may have to integrate with a blockchain through the APIs offered by the blockchain platform. When the protocols and hardware used by the existing supply chain systems are not compatible with the blockchain's API, it becomes challenging to interconnect a supply chain system with blockchain. While solutions such as Trinity, FoodTrust, and TradeLens provide a way to connect blockchain with the supply chain, it is still essential to develop interoperable APIs and protocols for the broader adoption of blockchain in the supply chain.

\section{Trusted authentication and access control for supply chain}
Supply chain applications and their management demand appropriate security in terms of access control and trust. In this section, we discuss the importance of access control and trust in the supply chain, and explore some available proposals. 

\subsection{Managing Access Control in Supply Chains}
Multiple organizations collectively provide and consume information in a supply chain network. Within each organization, a number of employees and devices may be responsible for registering supply chain data. Enabling access to the right entity with the right permission level is critical for the security of the supply chain. Note that a security breach at one organization could compromise the integrity of the entire supply chain. That said, the importance of access control in the supply chain is paramount. Access control is a security mechanism that assures the safe access of resources by the authorized objects. Access control places selective restrictions to access certain resources only by the trusted and authorised entities. These restrictions are governed by a set of access control policies that are controlled by an organization (or a number of organizations) that regulate access~\cite{pal2018policy}.




The access control mechanisms in the supply chain should take into consideration the various characteristics of a supply chain (e.g., distributed, time-sensitive, data-centric, etc) system from a multi-stakeholder point of view. Traditional access control mechanisms e.g., Role-Based Acess Control (RBAC), Attribute-Based Access Contro (ABAC), Organization-Based Access Control (OrBAC), and Capability-Based Access Control (CapBAC) cannot, in isolation, provide a fine-grained and robust access control solution for the supply chain~\cite{hussein2017community}~\cite{pal2021internet}. In RBAC, access is enforced based on the specific roles of an entity. In other words, in RBAC, access is granted for a certain resource to a specific role, and the users who belong to this role can then access the resource. However, RBAC is highly centralised, therefore, given the distributed and collaborative nature of a supply chain, it is not an ideal solution at scale. 

Alternatively, ABAC uses attributes (e.g., date, time, location, etc.) for enforcing access control policies, conditions and regulations. This approach provides much more flexibility in managing the identities of entities. However, ABAC does not use the concrete unique identity of an entity. This feature of ABAC may be promising for access control in the supply chain since it could provide some level of anonymity for the stakeholders. However, ABAC systems are also centralised in nature. Moreover, when the number of entities grows within the supply chain, ABAC by itself cannot handle the access control issues at scale~\cite{pal2018fine}. 

OrBAC uses entities (e.g., subject, action, and object) to control access over the resources. It allows the policy designer to explicitly mention security policies independent of the implementation. However, access control enforcement in OrBAC is highly centralised in nature, thus reduces commercial flexibility. CapBAC provides access control solutions based on capabilities. A capability can be defined as a communicable, unforgeable token of authority. It contains access control policies (and associated conditions) for specific resources. These capability tokens can be evaluated at the edge nodes at the time of access. However, given the scale and requirements of a supply chain network, managing the number of capabilities and their distribution is difficult~\cite{pal2017towards}. 

Access control for supply chain systems should be scalable, flexible, usable, trustworthy, and recognise the inherently decentralised nature of such systems. At the same time, the access control mechanism must be sufficient to protect the privacy, integrity, and confidentiality of the supply chain and its components~\cite{pal2019integration}. The issues noted above around the existing approaches to access control in the supply chain mean that further consideration of access control for the area is needed. Furthermore, the traditional access control models are user-centric and therefore they do not consider the relationship among various users, services, companies, and organisations. As noted earlier (cf. Section~\ref{introduction}), supply chain applications typically involve multiple stakeholders like producers, transporters, auditors, retailers, and regulators. Therefore, it is important to build trust between these entities for a scalable, flexible, and robust access control for the global supply chain network~\cite{zavolokina2020designing}.



We note that the use of blockchain in the supply chain attracted the interest of stakeholders. Blockchain technology is distributed and does not depend upon a centralised trusted authority - this minimizes the amount of trust required from a unit node in the blockchain ledger. To provide trust by commonly used access control mechanisms, in general, a centralised entity or a trusted third party acts as the intermediary to guarantee authenticity. In the case of blockchain, every node in the network is equal and ensures transaction authenticity by using consensus algorithms. So blockchain has the ability to provide a robust and flexible access control by providing distributed, secure, and trust-less features, which cannot be achieved using traditional access control solutions discussed above~\cite{song2021supply}. 

From a supply chain point of view, the core business and the sub-processes must create a trusted environment for a long-term partnership. That is, to increase customer value and ensure sustainable collaboration between the supply chain parties. However, this collaborative relationship needs a foundation of trust among the various entities within the supply chain network~\cite{kidd2003learning}.  Commonly, supply chain managers are not able to foster a trusting relationship among the various partners over the supply chain. Therefore, it is significant to incorporate different aspects of access control as well as other contexts (e.g., social, behavioral, etc.) that help to integrate a number of perspectives for the development of trust in products, suppliers, and customer relationships~\cite{sahay2003understanding}. Next, we discuss the significance of trust and its management in supply chain. 

\subsection{Trust Management in Supply Chain}

There is no coherent and universal definition of trust~\cite{pal2019towards}. It is a multi-dimensional and intangible concept without a clear understanding. This further creates difficulty in trust measurement, in particular, for large and complex supply chains. 
In different disciplines (e.g., social sciences and computing systems) the representation of trust is different. 

\begin{itemize}
    \item Trust in social sciences: It depends on social influences. This may come from characteristics and the individual behavior of an entity (e.g., a human) and are measured as honesty, cooperativeness as well as willingness to cooperate in a social setup. According to~\cite{samarati2000access}, trust can be seen as \textit{``the extent to which one party is willing to depend on somebody, or something, in a given situation with a feeling of relative security, even though negative consequences are possible''}.
    
    \item Trust in computing systems: It is different in different areas. That said, in computing the definition of trust differs depending on the domain e.g., networking, security, artificial intelligence, e-commerce, etc. According to~\cite{cho2010survey}, it is a subjective belief, as follows: \textit{``an agent’s trust is a subjective belief about whether another entity will exhibit behaviour reliably in a particular context with potential risks. The agent can make a decision based on learning from past experience to maximize its interest (or utility) and/or minimize risk''}. In~\cite{josang2007survey}, trust is defined as \textit{``the subjective probability by which an individual expects that another performs a given action on which its welfare depends''}.
\end{itemize}

Overall, trust can be referred to as the honesty, truthfulness, or even the reliability of a trustee. Trust is always context-dependent and seen the way it is used. From a supply chain point of view trust can be seen as: \textit{``at the heart of collaborative innovation capabilities. Without a foundation of trust, collaborative alliances can neither be built nor sustained. But, few companies are able to leverage trust for a sustainable competitive advantage; most companies lack high levels of trustworthy collaborations''}~\cite{fawcett2012supply}. 

Recall, supply chain applications are comprised of stakeholders that may situate in multiple countries and typically function in a distributed infrastructure (with internal partners, supplier, and buyer). In such a mobile and dynamic operational infrastructure, businesses are making decisions based on the data shared by different entities. Therefore the propagation of trust in the supply chain plays an important role in the logistic integration and collaboration of various components in the network to uphold the integrity of the supply chain~\cite{su2014research}. The organizations should not only trust the data values coming from different entities, but they should also rely on a trusted infrastructure. 

\begin{figure} [t]
    \centering
    \includegraphics[scale=0.45]{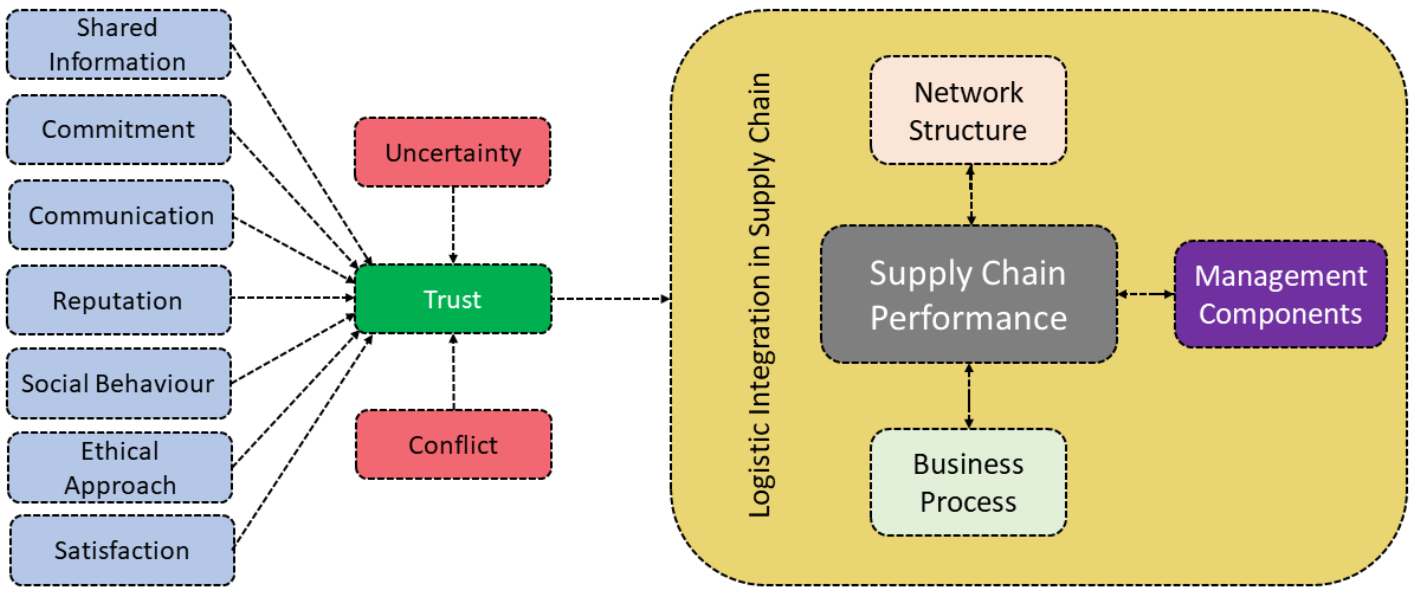}
    \caption{Propagation of trust and its integration in collaborative supply chain operation.}
    \label{fig:trust-ac}
\end{figure}

In Figure~\ref{fig:trust-ac}, we illustrate an outline of propagation of trust in the supply chain and its integration in collaborative supply chain operation. On the left side of the figure, we show various antecedents of trust for the supply chain. They are: (1) \textit{shared information} among the different entities that help to build up mutual expectations, (2) \textit{commitment} to the long-term orientation with both parties cooperating in supply chain values to maintain a trusted relationship, (3) \textit{communication} must be efficient and effective and should allow for feedback, (4) \textit{reputation} is an important factor to building trust by monitoring suppliers (and other components) to evaluate the performance, (5) \textit{social behavior} effectively highlights the issues of cooperation trust for making awareness and sensitivity on social and environmental issues, (6) \textit{ethical approach} represents corporate social responsibility to produce products and services, and (7) \textit{satisfaction} helps to improve on-time service, production, and delivery to the supply-chain members~\cite{kac2016influence}. 

However, the integration of these trust `components' is highly influenced by the \textit{uncertainty} and \textit{conflict} present in the environment. Uncertainty may come from unpredictable events, for example, fluctuations in supply requirements,  malicious activities, and frequent modifications to the parts of a supply chain management system. Similarly, conflicts may arise due to disagreements of opinions, business processes, and goals between the buyers and suppliers. Trust in the supply chain has a greater impact on the integration and management of the business process, management components, and network structure for improving operational performance~\cite{kolluru2001security}. Blockchain plays a significant role in this case - in the aggregation of different trust values (can be seen as the measurement of trust from various entities) that help the stakeholders validate the quality of information in supply chain~\cite{hou2014effects}. 

\cite{chen2017blockchain} discuss a trust-based supply chain quality management framework supported by blockchain technology. The discussion is focused on the various use-cases to build a trust management framework within a supply chain network. With a similar view, proposal~\cite{agrawal2021blockchain} discusses the a blockchain-based traceability framework for traceability in multi-level supply chain in textile and clothing industries.  \cite{al2021blockchain}, discuss the integration of blockchain technology in the supply chain to address trust issues and how to preserve data integrity between supply chain parties. The proposal also focuses on the efficient transmission of information between the partners in the supply chain. Blockchain is used to verify logged data and reproduces the actual observation of the entities within the supply chain. It helps in trust establishment for the actual (i.e., original) data by ensuring that the blockchain-logged data will be considered trustworthy.

\cite{shahid2020blockchain} discuss a trust management framework for agricultural supply chain supported by blockchain technology. In this model, every transaction is written inside the blockchain which finally uploads the data to a Interplanetary File Storage System (IPFS) - which is a protocol and peer-to-peer network that store and share data. The storage system then returns a hash of the data (stored inside the blockchain) to ensure trusted cooperation between the entities in the supply chain network. Similarly, \cite{bai2021blockchain}, discuss a trust management framework to manage the equipment in agricultural supply where the trust values of sensors are stored in the blockchain. \cite{malik2019trustchain} present a trust management framework, called `TrustChain', for supply chain management using blockchain technology. The proposed framework uses a reputation model that evaluates the quality of commodities as well as the trustworthiness of entities based on multiple observations of supply chain events. The reputation score comes from various segments of a supply chain network - separate from participants and products. Blockchain helps in a transparent, efficient, secure, and automated calculation of reputation score to build trust in the supply chain. In~\cite{longo2019blockchain}, a software connector is developed to connect a blockchain with the enterprises' information systems. This proposal aims to allow various companies to share information with one another with different levels of visibility. This in turn, build the trust by  checking data authenticity, integrity and invariability over time through the blockchain. 

\subsection{Section Summary}

In summary, access control and trust are two significant components for building a secure, safe and reliable supply chain infrastructure. Access control preserves confidentiality, integrity, and availability, and trust helps to maintain the integrity of information and relation among the various components. Access control, therefore, helps in trust-building in a supply chain infrastructure~\cite{pal2017design}~\cite{rabehaja2019design}. Typically, trust can be observed as a metric that is gathered by the interactions and observations based on the actors involved in a supply chain infrastructure. Finally, blockchain integration in the supply chain can provide secure, scalable, and interoperable communication of information with their trading partners with better access control management and addressing trust issues.

\section{Interoperability}
Supply chain networks are increasingly decentralised and global. The distributed, heterogeneous and regulated characteristics of supply chain demand blockchain systems to have inherent features for interoperability. Interoperability can be defined as the ability to communicate and access information across various blockchain systems. As there may be multiple small scale solutions and use-cases within supply chains, interoperability of these solutions is important for compliance and data sharing purposes. \par

Interoperability is needed as blockchain solutions within supply chain co-exist on enterprise level. These solutions differ depending upon the mutual interest of stakeholders, for example, technology and platform choices, commercial sensitivity of the data, governance and access control, scalability, etc. 
Furthermore, multiple blockchain ecosystems can be utilised together in a supply chain for various services and functionalities, e.g., a public blockchain can be used for digital payments, whereas a separate consortium blockchain can be used to record supply chain transactions and provide  identity and trust management services as in \cite{malik2021tradechain}.
 These blockchain solutions must be able to \textit{communicate} seamlessly without having to worry about the technical and design differences within each ecosystem. To enable these independent blockchain solutions to effectively communicate, careful categorization of the interoperability challenges is required.

\subsection{Interoperability Challenges}
In this section, we will review the main challenges with respect to interoperability of blockchain enabled supply chain systems.
\begin{itemize}
    \item \textbf{Governance and Data Privacy} As discussed in Section~\ref{characteristics-requirements}, one of the significant requirements of supply chain applications is the regulatory compliance that varies according to the country and different legal jurisdictions. More importantly, auditing and regulation heavily relies on the stakeholders and their willingness to share data across global boundaries. Due to the obvious competitive advantage, data privacy remains of utmost importance which may lead to interoperability barriers. Thus, lack of standardization and governance is a challenge for traceability as well as interoperability.
    \item \textbf{Platform and Data Heterogeneity:} Blockchain platforms are heterogeneous in  nature even within the use-case of supply chain systems. There must exist some technical compatibility among two blockchain platforms to communicate with each other in terms of consensus mechanisms, smart contract operations and data format. Furthermore, blockchain systems can store various types of data, for example, raw data, hashed event data, and encrypted data. Thus, in addition to blockchain platform heterogeneity, data heterogeneity also introduces challenges for interoperability. 
    \item \textbf{Blockchain Infrastructure:} The components that enable blockchain services may include the network infrastructure, back-end oracles, network nodes, cloud servers, etc.  Enabling disparate blockchain systems with propriety and legacy enterprise systems to communicate and work seamlessly with each other is another interoperability challenge.

\end{itemize}
    
\subsection{Existing Approaches for Interoperability}
Blockchain interoperability has been a topic of recent interest in literature. Most of these approaches target the interoperability challenge in general, without focusing on the application specific blockchains such as in supply chain or healthcare. In this section, we first discuss some of the literature in interoperability of blockchains and then categorise these approaches towards the end. \par 
 \cite{lafourcade2020blockchain} discuss the possibility of two blockchains to be inter-operable. According to their analysis, two public blockchains are only inter-operable if they follow the same transaction, consensus and block structure which is conceptually equal to having a single blockchain. Similarly, \cite{hardjono2018towards}, present a design philosophy for interoperability of blockchains.
 Blockchain interoperability can be categorised into mechanical and value levels. Mechanical level involves protocols, encryption standards, consensus mechanisms, transaction structure, etc. whereas value level corresponds to real assets, fiat currencies, etc. which can be associated with coins or tokens. The authors highlight that interoperability at the mechanical level is necessary for interoperability at the value level but it may not guarantee it. In their subsequent work \cite{hardjono2019toward}, they present blockchain gateways as a key notion of interoperability between two blockchains. Dedicated gateways within each blockchain system interact with each other for an asset transfer from one blockchain to another.\par
 A distributed publish/subscribe model, Trinity, with blockchain-based immutability is presented by \cite{8751388}. Instead of using a centralized broker, the authors propose a blockchain enabled pub-sub broker system. The data to be exchanged is distributed among all the brokers in the network through a consensus mechanism, validated through smart contracts and then stored on the ledger. Trinity evaluations exhibit that it consumes minimal resources, and the data management processes can be automated using the smart
contracts.\par
\cite{ghaemi2021pub} present an interoperability solution for permissioned blockchains using a similar publish/subscribe architecture. The proposed solution uses a broker blockchain network as opposed to any third-party. Broker blockchain uses its connector and topic smart contracts to provide connectivity among two other types of blockchains: a \textit{publisher} blockchain network and a \textit{subscriber} blockchain network. Publisher blockchain is the source from which the information is required by a subscriber blockchain. The broker blockchain stores the \textit{topic}, a copy of the information that needs to be shared between the two blockchain networks. The publisher blockchain not only creates a topic on broker blockchain but is also responsible to update any change in the information constituting to the topic. A PoC was developed using Hyperledger Besu, an Ethereum client, and two different versions of Hyperledger Fabric.\par

An interoperability architecture between private and public blockchain platforms is proposed by \cite{ghosh2021leveraging}. The proposed solution, \textit{CollabFed}, leverages on decentralised gateways and smart contracts. A consumer request for information is generated from an open network and logged on the public blockchain network. The consortium members of the private blockchain also take part in endorsing these user requests by a two-third majority vote. The logged requests trigger the smart contracts in private blockchains which propagate and schedule the requests based on a predefined business logic. The results are then transferred back to the consumers. PoC implementation of CollabFed was developed using Ethereum as the public blockchain platform and Hyperledger Fabric, and Burrow as the private blockchain platforms.\par
HyperService is a platform proposed by \cite{liu2019hyperservice} that delivers interoperability and programmability across
heterogeneous blockchains. HyperService contributes a programming framework that allows developers to build cross-chain applications in a unified programming model. In addition, it also provides the cryptography protocol that enables realising these cross-chain applications on blockchains.\par
An interoperability API implementation, called Bifrost is presented by \cite{scheid2019bifrost}. Bifrost allows users to store and retrieve arbitrary data on multiple blockchain systems. Bifrost consists of three components: APIs, blockchain adapters and a database. The API consists of store, migrate and retrieve function calls. The blockchain adapters covert user data into a transaction or a query based on store or retrieve function calls. Each blockchain must possess an adapter for communicating with other blockchains. Finally, the database stores the transaction hashes and necessary credentials required to retrieve data. \par
A token based cross blockchain platform for blockchain interoperability is presented by \cite{borkowski2018caught, borkowski2018deterministic}. A cross blockchain asset transfer token, PAN, is introduced using claim-first transactions. The authors describe a method of generating a cryptographic Proof of Intent (PoI). PoI certifies that the sender on source blockchain system is willing to transfer a given amount of assets to a wallet address 
on a destination blockchain. This PoI then can be used on the destination blockchain to claim the transferred assets. To update the asset records on source blockchain, observing parties (called witnesses) are incentivised for their role.\par
Various other literature outline the need and requirements of blockchain interoperability.  A good survey on blockchain interoperability is provided by \cite{johnson2019sidechains} and \cite{ belchior2020survey}. To summarize, the interoperability approaches can be categorised into the following broad categories:

\begin{itemize}
    \item\textbf{{Side chains, Management chains and Relays}}: Side chains are considered to be the blockchains which are typically the extension of the main chain. A cross-chain communication protocol is then used for asset exchange between the two chains. Sometimes they are also used as relays to incorporate offline payment or query mechanisms from a financial chain to the main chain. A few examples of such interoperability mechanisms are Broker blockchain by \cite{ghaemi2021pub}, Polkadot \cite{ wood2016polkadot}, Ethereum 2.0 Sharding with beacon chain, AION, Blocknet, etc.
    \item\textbf{{Notaries}}: A notary is a trusted entity or an organization who acts as an intermediary between the two blockchain systems and monitors transactions and triggers. It is responsible for exchange of information from the source to destination blockchain. These schemes are employed as centralised or decentralised exchanges, such as Uniswap by \cite{adams2021uniswap}, and Liquid by \cite{blockstream}.
    
    \item\textbf{{APIs/Gateways}}: An application programming interface (API) is a piece of code that governs the access point to a server and the rules developers must follow to interact with a database, library, software tool or programming language. APIs or gateways act as a translator, taking requests from source blockchain and converting them to language or format understandable by the destination blockchain. Application of such approaches are discussed above in \cite{liu2019hyperservice} and Bifrost by \cite{scheid2019bifrost}. A very well known application is Hyperledger Cactus, a blockchain integration framework which validates cross chain transactions using a validator network.
    
    \item\textbf{{Hashed/Time locked contracts (HTLCs)}}: These approaches use hashlocks and time locks to enforce operational atomicity between the two parties, typically on disparate blockchain systems. A trader provides a cryptographic proof of committing to a transaction before a time-out. HTLCs enable cross-chain atomic operations or atomic swaps, such as for conditional payments from Bitcoin to Ethereum. Few other examples include Bitcoin lightening network and BTC relay.
\item {\textbf{Pub-Sub Models and Tokens:}}  Pub-Sub models employ the concept of message exchange between publisher and subscriber blockchain networks. Publisher blockchains are often termed as the source of information whereas the subscriber is the requester. The message exchange can be done through either tokens, APIs or side-chains such as systems proposed by \cite{8751388}, \cite{ghaemi2021pub} and \cite{borkowski2018caught, borkowski2018deterministic}.
\end{itemize}

\subsection{Suitability of Interoperability Approaches in Supply Chains}
The approaches discussed in the previous section can widely be adopted depending upon the interoperability requirement. In supply chains, governance and regulatory compliance is very important for monitoring and auditing purposes, which can be achieved  by adopting an interoperability approach based on side chains combined with notaries.
A consortium of notaries may play an important role in defining access policies for privacy sensitive data. For integrating financial transactions with physical asset exchanges, tokenization and Pub-Sub models are best suited. Similarly, the APIs and gateways may help to reduce the infrastructure challenges by translating data read/write requests. Although HTLCs and smart contract approaches are complex in terms of implementation and may not be generalised in their design, these approaches are well suited for supply chain systems to enforce standardization based on legal jurisdictions across various chains.  
\section{Importance of Scalable Blockchains for Supply Chain}
Supply chain involves multiple stakeholders that communicate through transactions and smart contracts leading to a huge volume of transactions in blockchain. Conventional blockchains are not directly applicable in supply chain as they suffer from low scalability which is rooted by blockchain resource consumption, latency,  efficiency and  throughput which are discussed in greater details below.
\begin{itemize}
    \item \textbf{Resource Consumption:} Recall that blockchains are managed distributively by all participating nodes. The conventional methods that facilitate the distributed management of blockchain incur significant computation, memory, and bandwidth overheads which limits its scalability. The computational overhead of the blockchains is rooted by    resource consuming consensus algorithms. Additionally,  Once a new transaction/block is generated, all participants shall verify the same. The verification of transactions also involves verifying the history of transactions which requires the verifies to store blockchain database. However, the size of the blockchain database will significantly increase as due to its immutability, removing or modifying previously stored data is not possible and will compromise the blockchain consistency. Blockchain broadcast the blocks/transactions which consumes significant bandwidth due to large number of participants in supply chain applications.
    
    \item \textbf{Latency:} There is a non-trivial delay associated with committing a new transaction in the blockchain and receiving confirmation. This delay involves the delay in committing the transaction (i.e., following the consensus algorithm) and the delay in receiving confirmation. The latter involves waiting for a particular number of blocks to be appended to the block in which a particular transaction is stored  which in turn protects against double spending and ensures the transaction will not be placed in forked blocks. However, as shown in Figure \ref{fig:overview}, supply chain involves transactions that require real-time transaction settlement. 
    
    \item \textbf{Efficiency:} In most of the existing consensus algorithms, multiple validators, i.e, miners, work simultaneously to commit the same block in the blockchain and only the one that follows the consensus algorithm first is permitted to commit the block and thus receive incentive while the resources of other nodes will be wasted. This in turn limits the blockchain efficiency.

    \item \textbf{Throughput:} Conventional blockchains suffer from limited throughput, i.e., total number of transactions that can be committed in blockchain per second. However, supply chain applications demand high throughput due to high transaction generation rate. The number of transactions is continuously increasing as new applications and users join the system which highlights the demand for a throughput management algorithm. The latter ensures that the blockchain throughput can accommodate the load in the network.
    
\end{itemize}
 
Blockchain optimization for applications beyond cryptocurrency, including supply chain, has received significant attention in recent years. The authors in \cite{thakur2020scalable} employed off-chain transactions to increase the blockchain scalability. The participating nodes that are involved in the life time of a product, jointly create a channel where all transactions related to the product are stored. This in turn reduces the number of transactions that needs to be committed in the blockchain as for each channel only two transactions will be committed in the blockchain. The rest of the communication happens in the channel. \par 

To increase the blockchain scalability, the authors in \cite{malik2018productchain} employed the concept of \textit{sharding}. In general, sharding refers to dividing the network into smaller groups, i.e., shards where the transactions/blocks generated in each shard are broadcast and verified inside the shard. Thus, sharding increases the blockchain scalability by limiting the scope of the ledger. Sharding improves the blockchain throughput and reduces delay linearly as the improvement rate is directly impacted by the number of shards that exist in the network. \par

The authors in \cite{dorri2021vericom} introduced a verification and communication model for blockchain, known as \textit{Vericom}, to  the blockchain bandwidth consumption by introducing a dynamic multicasting and traffic routing algorithms. To route traffic, i.e., blocks and transactions, the authors employed a PK-based routing algorithm where the traffic is routed based on the PK of the destination. A group of high resource available nodes in the network jointly form a backbone network.  Each backbone  node is allocated to a particular \textit{routing character} which refers to the most significant characters of the hash function output. Depending to the hash of their PK, the participating nodes join a backbone node to receive transactions sent for them. The backbone nodes use conventional IP-based routing algorithms to route traffic in backbone. Unlike conventional blockchains where transactions are broadcast, Vericom multicasts the transactions to a randomly selected group of nodes known as verifier sets. The latter is unique per transaction/block and is selected based on the hash function output of the transaction/block content. \par

The consensus algorithm is  fundamental in blockchain scalability. Most of the conventional consensus algorithms consume significant resources, limit throughput, lack efficiency,  and involve significant delay in committing and confirming transactions. Various optimized consensus algorithms have been introduced in recent years \cite{sankar2017survey}. Intel introduced Proof of Elapsed Time (POET) \cite{sankar2017survey} consensus algorithm which relies on the Trusted Execution Environment (TEE)  in Intel CPUs. When a validator aims to commit a new block  to the blockchain, it needs to wait for a random period of time identified by TEE. In case during the waiting period, it receives a block consisting the same transactions, it shall drop the block and start committing a new block. \par
 
In \cite{dorri2020tree} the authors introduced a scalable fast consensus algorithm with near real-time transaction settlement which is known as \textit{Tree-chain}. Tree-chain is a leader-selection consensus algorithm where a leader is selected for a particular period of time and commit transactions, with specific features, to the blockchain. Tree-chain relies on the hash function output to achieve randomization  in two levels: blockchain level and transaction level. In blockchain level, a \textit{Consensus Code Range (CCR)} is distributed randomly and in an unpredictable way between the validators. CCR refers to the most significant characters of the hash function output. Each potential character in the hash function output is associated with a weight defined in a weight dictionary. The validators calculate a Key Weight Metric (KWM) which is essentially the sum of the weight of all the characters in the hash of a PK. The KWMs are then stored in a list in descending order and the node with the highest value of KWM is dedicated to the first CCR range.\par
In transaction level, the validator of each transaction is randomly selected based on the hash of the transaction content. The validator whose CCR matches with the most significant characters of the hash of a particular transaction, is the validator in charge for committing that particular transaction in the blockchain. Each validator creates a unique ledger chained to the genesis block for committing its corresponding blocks which ensures fast transaction settlement. Thus, Tree-chain, by design, embraces the concept of forking. \par
\section{Design Considerations and Open Challenges}
This section reviews the design considerations and open challenges to help stakeholders carefully adopt blockchain technology for their supply chain applications.

\subsection{Public vs. Permissioned Blockchains}
A blockchain-based supply chain application can be designed using either a permissioned or a public blockchain. When selecting a blockchain platform, the stakeholders must carefully weigh the pros and cons of these approaches. 

A public blockchain primarily relies on nodes maintained by the community members. In the case of BitCoin[\cite{nakamoto2019bitcoin}] and Ethereum[\cite{,wood2014ethereum}], any community member with a computationally capable node can join the network and participate in the PoW consensus process. As discussed in Section~\ref{sec:blockchain}, the node that successfully solves the cryptographic puzzle gets rewarded with a cryptocurrency. Here, the reward comes from the users that submitting transactions to the network. It is important to note that the user must pay a transaction fee whenever a transaction is submitted to the blockchain network. Therefore, applications that rely on public blockchain platforms must consider the transaction fees. Note that a supply chain application that produces hundreds of transactions per day must spend tens to hundreds of dollars per day for leveraging the services of a public blockchain. 

Alternatively, applications can use a permissioned or private blockchain for a supply chain application. Following this model, the stakeholders that are part of a supply chain network would share their computing nodes for managing and maintaining the blockchain in a distributed and trusted fashion. Technically, a private blockchain network can be established with platforms such as Hyperledger Fabric[\cite{androulaki2018hyperledger}], Tendermint[\cite{buchman2016tendermint,androulaki2018hyperledger}], or a customized private version of Ethereum. Such platforms typically employ a lightweight consensus algorithm, which differs from a computationally expensive PoW consensus model. Since the blockchain network is established among the stakeholders that are part of a supply chain network relevant to their business following a lightweight consensus process, users are not required to pay a transaction fee. It is essential to note the operational cost of running a private blockchain network are covered by businesses that leverage them. Besides, the cost would be marginal since the companies are using the Internet and computing code without relying on computationally intensive consensus processes. And, the blockchain itself does not demand users to pay a transaction fee, as seen in public blockchains. Therefore, private blockchain platforms are cost-effective while allowing organizations to leverage trust and immutability in a distributed and multi-stakeholder settings.

In summary, the stakeholders must weigh in these pros and cons when selecting a blockchain platform as the operational cost of a blockchain may discourage some stakeholders from submitting transactions to a public network, potentially reducing the effectiveness of a blockchain-based supply chain application [\cite{DEGIOVANNI2020107855}].

\begin{figure}
    \centering
    \includegraphics[scale=0.35]{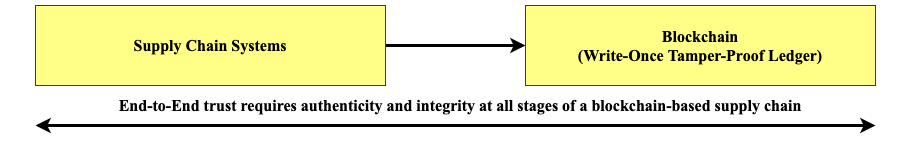}
    \caption{Garbage-In-Garbage-Out problem: Storing data in a blockchain ledger alone does not guarantee authenticity and trust. All the layers of a blockchain-based supply chain requires trusted methods.}
    \label{fig:gigo}
\end{figure}

\subsection{Preventing Garbage-In-Garbage-Out Problem}
Blockchain technology allows the stakeholders to track and trace their products throughout the supply chain reliably. When products and goods move through a supply chain, each stakeholder must gather and log data associated with them. For example, the arrival time, temperature, humidity, and expiry date of a product must be logged at each supply chain endpoint to ensure that the food products are in edible condition. Under this operational setting, blockchain technology can receive information about the products and store them in an immutable ledger following a consensus process involving multiple stakeholders in a distributed network. It is essential to underscore how blockchain can also receive incorrect information from the stakeholders and store them in a blockchain. Note that blockchain technology does not offer any mechanism to verify the correctness of the data submitted by a user. This issue is referred to as \emph{Garbage-In-Garbage-Out} problem[\cite{ziolkowski2020decision}], which means the blockchain technology can also store inaccurate information in the ledger and present them to the users in the future. Stakeholders may think that blockchain technology is helping them provide transparency to the end-users. On the contrary, inaccurate provenance information may mislead the end consumers while ruining the reputation of the organizations that are part of the supply chain network. Therefore, the stakeholders must ensure that the blockchain ledger's information is accurate for a genuinely dependable supply chain ecosystem. Figure~\ref{fig:gigo} highlights the importance of end-to-end trust.

Garbage-In-Garbage-Out problems require solutions outside of the blockchain. In particular, the interface between the physical and the digital world must be built using robust methods to ensure end-to-end trust and transparency. Therefore, supply chain applications must include approaches involving humans, sensing devices, trusted third parties including regulatory bodies and certification laboratories to weed out false claims in a multi-stakeholder supply chain system.

\subsection{Automated Verification of Compliance}
\label{sec:gigo}
A supply chain network is established involving multiple stakeholders operating from numerous jurisdictions. Within each jurisdiction, stakeholders must comply with the regulations enforced by the local government and other regulatory bodies. In a supply chain network, examples of regulations include export controls and technical regulations. Contemporary supply chain applications rely on paperwork, and many human operators at various jurisdictions ensure regulatory compliance. When leveraging blockchain technology for the supply chain, it may be easier to automate the compliance verification process through a smart contract[\cite{7987376}]. Recall that smart contracts allow the stakeholders to track and manage the supply chain autonomously. Authorities could translate the regulatory standards into one or more smart contracts to speed up the compliance verification process.

It is important to note that the automated verification of regulatory compliance requires accurate data from the stakeholders. For regulatory authorities to adopt automation in this context, the Garbage-In-Garbage-Out problem discussed in Section~\ref{sec:gigo} must be solved. 

\subsection{Lack of Common Data Standard}
When establishing a supply chain network involving multiple stakeholders from multiple jurisdictions, each stakeholder may follow a custom data standard local to their country. An absence of a common data standard in a distributed and multi-stakeholder setting would result in interoperability issues, wherein each stakeholder may have to develop new extensions to their existing infrastructure, which would hamper adoption [\cite{renner1996data}]. It is important to note that an organization responsible for shipping products may be dealing with several products belonging to various stakeholders. For such an organization, it is almost impossible to develop new extensions to interface with each organization seamlessly. 

Besides, supply chain applications may want to store a combination of media files such as images and videos and digital data to enable provenance. Blockchain technology can handle a few bytes of data, but it is not a good fit for storing large data items such as videos and images. In such cases, stakeholders may have to keep the data either locally and store the hash on the blockchain or in decentralized file storage such as IPFS [\cite{benet2014ipfs}] and store the file pointer on the blockchain, as discussed here [\cite{8656952}].

Given these issues around the absence of a common data standard and large data files, supply chain applications must develop and adopt a universal and application-agnostic data standard.

\subsection{Privacy Concerns}
The information about the products helps the stakeholders understand the status of the supply chain. Depends on the product and the supply chain network, the stakeholder may have to share sensitive details, including the ingredients used (in the case of a food producer) and location (in the case of shipping companies). Stakeholders may not be willing to openly share sensitive details since it may leak their intellectual property or other business secrets. In transportation and shipping companies, people's location information gets exposed to several stakeholders. Stenberg \textit{et al.}[\cite{https://doi.org/10.1111/jbl.12240}] reports that the truck drivers are not comfortable with live tracking of their location in a real-world supply chain. Besides, the data protection regulations [\cite{pantlin2018supply} such as General Data Protection Regulation (GDPR), California Consumer Privacy Act (CCPA), and Australian Privacy Act (APA) provide guidelines about the use of data associated with products and companies operating from a particular jurisdiction. Such regulations and the privacy concerns of the stakeholders must be considered when designing a blockchain-based supply chain application. Malik \textit{et al.} introduce PrivChain[\cite{malik2021privchain}], a blockchain-based solution for protecting the privacy of the supply chain stakeholders using Zero-knowledge proof and Pederson commitments.

\subsection{Lack of Interoperability with Legacy IT Systems}
Many supply chain organizations have been using enterprise systems to manage their supply chain processes. Such systems enable stakeholders to share information about their products, handle payments, and generate paperwork for regulatory compliance. Note that organizations may have made significant investments to build and deploy these systems[\cite{https://doi.org/10.1111/jbl.12240}]. Connecting these legacy systems with a blockchain-based supply chain may require a considerable overhaul, making it more expensive and disruptive. Therefore, blockchain-based supply chain applications must support protocols and services leveraged by legacy and enterprise systems. 

\subsection{Operational Costs}
Supply chain systems need a set of new processes and methods to establish a trusted blockchain-based supply chain. Such extensions may require personnel with specialized skills. Besides, organizations may also have to invest in new hardware and software infrastructure. Such issues may make organizations hesitate to adopt a new blockchain-based supply chain system. Allowing organizations to leverage legacy systems and processes would reduce the operational cost to a large extent. 

In addition, organizations have to deploy consensus nodes and manage them if they are part of a private blockchain network. The deployment and management of consensus nodes introduce a significant overhead both in terms of hardware and networking. IBM's TradeLens and FoodTrust offer blockchain-as-a-service to organizations. Following this model, companies need not deploy new hardware for leveraging blockchain for traceability and provenance. However, organizations must trust the third-party service provider since the blockchain infrastructure is not owned and managed by multiple organizations in a distributed and transparent setting. 


\subsection{Lack of Engagement from Field Operators}
A supply chain application involves field operators responsible for driving trucks or manually recording information at a supply chain terminal. The information entered by these field operators is highly critical for tracking the flow of goods in a supply chain. The integration of blockchain to a supply chain network may expose some of the data entered by these field operators to other entities in the supply chain in a raw format. Note that current systems may record the raw data in a server belonging to the field operator's organization. As part of the information sharing process, each organization may only share a summary or high-level statistics with other organizations, filtering sensitive information recorded by the field operators. With the introduction of blockchain, sensitive information may get stored directly on the blockchain, which may infringe the privacy of the field operators. Due to the perceived risk of privacy violation, field operators may hesitate to record data, which would reduce the effectiveness of blockchain-based supply chain applications. The integration of privacy-preserving features may provide confidence to the field operators, which would increase their engagement with the system.

In addition, blockchain-based supply chain systems should not include cumbersome processes demanding additional effort from field workers. For example, a truck driver may not favor logging into a terminal every few minutes to register the status of the supply chain [\cite{https://doi.org/10.1111/jbl.12240}]. Therefore, it is essential to consider how the new processes affect the day-to-day work of the field operators.

\begin{figure}
    \centering
    \includegraphics[scale=0.4]{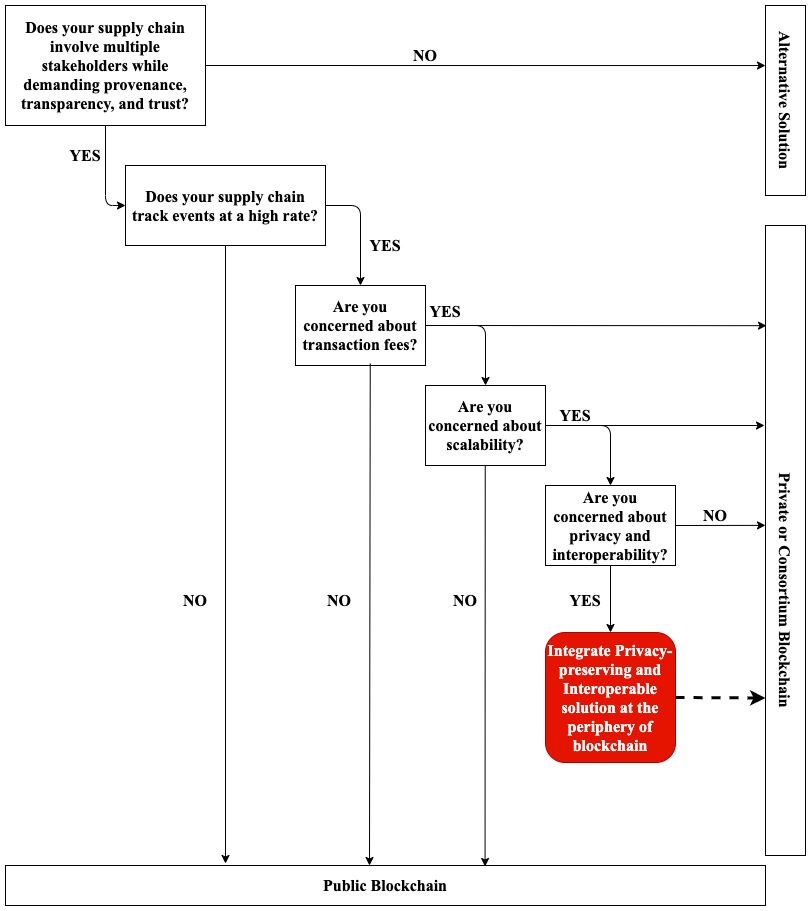}
    \caption{A Decision-tree for Selecting the Suitable Solution for Supply Chain Applications}
    \label{fig:decision-tree}
\end{figure}

\subsection{Payment Processing Challenges}
In a supply chain application, multiple stakeholders provide products and services to other stakeholders. When a physical product moves in a forward direction towards the end consumers, the payment associated with products flows in the reverse direction. The literature on supply chain argues the need for the inclusion of a digital infrastructure and automation processes to prevent payment delays [\cite{viriyasitavat2021augmenting}]. Having a cumbersome manual process with paper-based documents would be detrimental to supply chain efficiency. Therefore, digitization and the automation of the payment process would be massively beneficial to the supply chains.

Blockchain technology includes smart contracts, an innovative and decentralized computation engine. Using smart contracts, businesses can code up the payment process, which supply chain events could trigger. However, the events that drive the smart contracts must carry accurate information since the organizations may not want to make the payment without having confidence in the data. Therefore, the \emph{garbage-in-garbage-out} problem discussed in Section~\ref{sec:gigo} must be resolved to confidently automate payments.

\subsection{Sustainability Demands}
Global warming and ever-growing carbon emissions force authorities to impose stricter regulations around sustainability. Supply chain companies are asked to incorporate sustainable standards and practices in their operations. Many of the existing blockchain-based supply chain systems predominantly track the flow of products towards the end-consumers and process payments in the reverse direction. Recycling of products is not given much importance. The emerging sustainability standards emphasize the importance of recycling. Many blockchain-based solutions have been created to incentivize sustainable behavior in the food supply chain [\cite{chidepatil2020trash,zhang2019end}]. Following these initiatives, the supply chain stakeholders must introduce mechanisms to manage the recycling and decommissioning processes. Note that the blockchain technology offer support for tokens, which can be used to incentivize end consumers for their sustainable actions [\cite{patel2020carbon,saraji2021blockchain}].

\section{Decision Tree for Supply Chain}
Figure~\ref{fig:decision-tree} provides a guideline for supply chain stakeholders. It helps the stakeholders choose a suitable solution for their supply chain based on the design and operational constraints. In a nutshell, applications producing hundreds to thousands of events have to opt for a private blockchain to minimize the operational costs. From the cost perspective, the number of transactions is a good indicator, but other constrains including privacy and need for interoperability warrant new solutions at the periphery of the blockchain. While we acknowledge that this decision tree is not comprehensive for a supply chain domain, but it highlights the importance of minimizing the operational cost at the early design phases to establish an effective and sustainable supply chain.

\section{Chapter Summary}
Supply chain applications operate in a distributed and collaborative environment involving multiple stakeholders from various jurisdictions. Stakeholders including producers, regulatory bodies, logistics companies, and end consumers increasingly demand transparency and provenance to fully understand the history of products. Centralized digital and enterprise systems offer a viable solution for tracking products in a supply chain, but they suffer from single point of failure, wherein the organization that runs and manages the infrastructure may misbehave compromising the integrity of the entire supply chain. Blockchain technology offers features such as immutable ledger, decentralization, and smart contracts, which are tailor-made for the creation of a trusted and decentralized supply chain networks.

This chapter introduced the challenges and the requirements of a supply chain application. In particular, it introduced how blockchain technology can be used to share data in a multi-stakeholder supply chain environment. The discussions on trust and access control underscores the importance of security and authentication requirements. To help the supply chain stakeholders decide on the choice of blockchain platforms, this chapter discussed the scalability challenges of blockchain-based systems. The section on interoperability highlighted the need to support legacy systems and multi-blockchain frameworks to enable broad adoption.

Lastly, the section on open challenges and opportunities provided guidelines to supply chain stakeholders as they start to adopt blockchain technology for their supply chain. This chapter showed that blockchain technology offers several advantages to supply chain applications, but the issues around scalability, interoperability, privacy, regulations, and data trust must be solved to establish a dependable multi-stakeholder supply chain.


\bibliography{references}



\end{document}